# An energy-based perspective on the correlation between stress drop and rupture speed

**Shiqing Xu**[1]

[1]Department of Earth and Space Sciences, Southern University of Science and Technology, Shenzhen 518055, China

Correspondence to: S. Xu, xusq3@sustech.edu.cn

**Key Points:**

- Energy balance relation from fracture mechanics is applied to understand the correlation between stress drop and rupture speed
- Near constant fracture energy can yield a positive correlation between stress drop and rupture speed
- Variable fracture energy dependent on space, time or off-fault damage may cause a negative correlation between stress drop and rupture speed

## Abstract

Stress drop $\Delta\tau$ and rupture speed $V_{\mathrm{r}}$ are two important earthquake source parameters that control the characteristics of rupture process and the associated ground motion. However, how the two parameters correlate with one another is currently still under debate. Here I use an energy-based approach from fracture mechanics to understand the correlation between $\Delta\tau$ and $V_{\mathrm{r}}$. This approach is built on the balance between fracture energy $G_{\mathrm{c}}$ and dynamic energy release rate $G_{\mathrm{d}}$ (which itself is a function of $\Delta\tau$ and $V_{\mathrm{r}}$). By synthesizing various observations within a unified fracture-energy ($G_{\mathrm{c}}$) framework, I show that near constant $G_{\mathrm{c}}$ can yield a positive correlation between $\Delta\tau$ and $V_{\mathrm{r}}$, whereas variable $G_{\mathrm{c}}$ dependent on space, time or off-fault damage may cause a negative correlation between $\Delta\tau$ and $V_{\mathrm{r}}$. These results reconcile the debate on the correlation between $\Delta\tau$ and $V_{\mathrm{r}}$ and suggest a need to examine the condition of other factors (such as $G_{\mathrm{c}}$). Additional factors may complicate the evaluation of the correlation between $\Delta\tau$ and $V_{\mathrm{r}}$ when rupture process is inferred from far-field observations, accompanied by strong spatiotemporal variation, or followed by additional phases, which should be investigated by future studies.

## 1. Introduction

Rupture speed $V_r$ describes how fast an earthquake rupture propagates along the fault, and stress drop $\Delta\tau$ dictates how much strain energy pre-stored in the surrounding media is released by an earthquake. These two parameters hold great importance for understanding the physical mechanism underlying earthquake rupture process (Das, 2007; Gao HY et al., 2012; Weng HH and Ampuero, 2022), earthquake source scaling relations (Kanamori and Anderson, 1975; Kanamori and Rivera, 2004; Ye LL et al., 2016), and the intensity and distribution of ground motion (Aagaard and Heaton, 2004; Hanks and McGuire, 1981; Kwiatek and Ben-Zion, 2016). While the basic roles of $V_r$ and $\Delta\tau$ in characterizing earthquake source and ground motion have been generally recognized, how the two parameters correlate with one another still remains a highly debated topic. Some studies show a positive correlation based on direct experimental observations (Okubo and Dieterich, 1984; Passelègue et al., 2013; Svetlizky et al., 2017; Xu SQ et al., 2018; Chen XF et al., 2021; Dong P et al., 2023), whereas others report a negative correlation based on ground motion variability (Causse and Song, 2015) and source inversion of natural earthquakes (Chounet et al., 2018; Shimmoto, 2022; Nazeri et al., 2025).

Before further investigation on the debated topic, it is noteworthy that the above two contrasting views may partly stem from different ways of computing source parameters (e.g., dynamic stress drop for most laboratory earthquakes vs. static stress drop typically for natural earthquakes), or different frameworks for examining the related correlation (e.g., fracture mechanics as in Svetlizky et al. (2017) vs. ground motion as in Causse and Song (2015)). Moreover, the view of negative correlation for some natural earthquakes (Chounet et al., 2018; Shimmoto, 2022; Nazeri et al., 2025) may be biased by model assumptions, estimation uncertainties and tradeoff effects, since results are indirectly inferred from far-field observations (Kanamori and Rivera, 2004; Abercrombie, 2021; Ye LL et al., 2016). For these reasons, it is crucial to reconcile the debate on the correlation between $\Delta\tau$ and $V_r$ in a coherent framework, and to preferably test the view(s) with direct, near-field observations.

In this opinion-type article, I use the energy balance relation from fracture mechanics to understand the correlation between $\Delta\tau$ and $V_\mathrm{r}$. The idea of invoking the energy balance relation is not new and has been applied in previous studies to understand either a positive (Svetlizky et al., 2017) or negative (Gabriel et al., 2013) correlation between $\Delta\tau$ and $V_\mathrm{r}$. What is new in this study is the synthesis of various observations within a unified fracture-energy framework. Especially, I construct three representative scenarios, supported by theoretical arguments and observational evidences (Table 1 and Figure 1), to show that both positive and negative correlations are possible and the exact outcome can depend on the condition of fracture energy.

## 2. Energy balance relation

According to fracture mechanics, the equation of motion for a rupture front is described by the balance between fracture energy $G_\mathrm{c}$ and dynamic energy release rate $G_\mathrm{d}(V_\mathrm{r}, \Delta\tau) = g(V_\mathrm{r}) \cdot G_\mathrm{s}(\Delta\tau)$ (Freund, 1998):

$$G_\mathrm{c} = G_\mathrm{d}(V_\mathrm{r}, \Delta\tau) = g(V_\mathrm{r}) \cdot G_\mathrm{s}(\Delta\tau) = g(V_\mathrm{r}) \cdot \left\{ \frac{1-\nu^2}{E} \cdot \left[ \int_{-L}^{L} \frac{\Delta\tau(x)}{\sqrt{\pi L}} \cdot \sqrt{\frac{L+x}{L-x}}\, dx \right]^2 \right\}. \tag{1}$$

In the above equation, $g(V_\mathrm{r})$ represents a universal function that monotonically decreases from one towards zero with rupture speed $V_\mathrm{r}$ in the subshear regime (when $V_\mathrm{r}$ stays below the shear wave speed $C_\mathrm{S}$), where $V_\mathrm{r}$ itself is measured locally at the rupture front and can vary with space $x$ and time $t$ (for brevity, I use the notation $V_\mathrm{r}$ instead of $V_\mathrm{r}(x,t)$). $G_\mathrm{s}(\Delta\tau)$ is called static energy release rate and can be expressed as a function of stress drop $\Delta\tau$, Poisson's ratio $\nu$ and Young's modulus $E$ (Freund, 1998). For now, I assume $\nu$ and $E$ are material constants and mainly focus on the dependence of $G_\mathrm{s}(\Delta\tau)$ on $\Delta\tau$ through a functional integral (note that $\Delta\tau$ itself can vary with space and/or time). For the purpose of demonstration, a simple case of mode-II rupture (with a half-length of $L$) embedded in an infinite space has been assumed in the rightmost part of Eq. (1), where $\Delta\tau(x)$ stands for the (time-independent) magnitude of stress drop at location $x$ and $\sqrt{(L+x)/(L-x)}$ represents the associated weighting factor for transferring stress to the rupture front at $x = L$ (Tada et al., 2000, eq. 5.10). If stress drop obeys a spatially uniform distribution (i.e., $\Delta\tau(x) = \Delta\tau_0$), then the term inside the bracket in Eq. (1), known as static stress intensity factor $K_\mathrm{s}$, will be reduced to the well-known result $K_\mathrm{s} = \Delta\tau_0 \cdot \sqrt{\pi L}$ (Tada et al., 2000, eq. 5.1).

Besides the above, several additional points deserve to be mentioned regarding the conditions and possible extensions of Eq. (1):

(i) It is assumed that fracture energy $G_c$ is dissipated near the rupture front, for both the classical case with on-fault, small-scale yielding (Freund, 1998) and the extended case with off-fault yielding (Andrews, 2005).

(ii) As a consequence of (i), the impact of $G_c$ on $V_r$ is local, allowing for an abrupt change in $V_r$ (Freund, 1998).

(iii) The impact of stress drop $\Delta\tau$ on $V_r$ is nonlocal, due to an integral effect as illustrated in Eq. (1) and other studies (Dunham et al., 2003; Bayart et al., 2018).

(iv) Despite the nonlocal impact of $\Delta\tau$ on $V_r$, the relative importance of stress drop $\Delta\tau(x)$ at a specific location $x$ can be enhanced (or reduced), if it has a large (or small) magnitude and operates close to (or far away from) the rupture front at $L$, due to the factor $1/\sqrt{L-x}$ in Eq. (1).

(v) So far the energy balance relation in Eq. (1) has been rigorously verified for mode-II crack-type ruptures in the subshear regime (Svetlizky et al., 2019, and references therein), but it can also be extended to pulse-type or mode-III ruptures, by replacing the length scale $L$ with the one related to pulse width or by adopting a different monotonic function for $g(V_r)$ (Rice et al., 2005).

(vi) It is not fully clear how Eq. (1) can be extended to the case with nonlocal fracture energy dissipation that takes place over a long distance behind the rupture front (Brener and Bouchbinder, 2021). Solving this problem requires careful consideration of local and global scales (Ben-Zion and Dresen, 2022; Kammer et al., 2024), and accurate evaluation of whether additional stress drops in the rupture tail region can still influence the rupture front (i.e., mind the domain of dependence and the region of influence in a wave/rupture propagation problem) (Ding XT et al., 2024).

(vii) If stress drop shows a complicated spatiotemporal distribution (Freund, 1998, chap. 7), if other scales emerge in the source region or surrounding media (Goldman et al., 2010; Buehler et al., 2003; Rice et al., 2005; Weng HH, 2025), or if symmetry is broken in the considered problem (Aldam et al., 2016), then the specific expression inside the bracket in Eq. (1) (which affects $K_s$ and $G_s$) will need to be modified.

(viii) If rupture expansion style changes (e.g., from spontaneous to self-similar) (Broberg, 1999), if the finiteness of space becomes important (Goldman et al., 2010; Weng HH and Ampuero, 2019), or if material asymmetry emerges (Rubin and Ampuero, 2007), then the detailed form of $g(V_r)$ in Eq. (1) will also need to be modified. Nonetheless, the general decreasing trend of $g(V_r)$ with increasing $V_r$ may still hold for a range of realistic ruptures, reflecting the basic feature of less available energy for advancing rupture front at faster $V_r$ (due to more partitioning for radiated energy) (Kanamori and Rivera, 2006).

In Section 3, I will use the energy balance relation in Eq. (1) and the above supplementary points (i–viii) to deduce specific correlations between $\Delta\tau$ and $V_r$, by adjusting the condition of $G_c$. It should be emphasized that the main goal is not to explore all the possibilities over a large set of scenarios, but to elucidate a positive or negative correlation in some selected scenarios. These selected scenarios together with the supporting observations are sufficient to confirm a variable correlation between $\Delta\tau$ and $V_r$. While Eq. (1) implies that an integral (nonlocal) form of $\Delta\tau$ and a local value of $V_r$ should be used to deduce correlations, in practice it could be difficult to accurately estimate the detailed $\Delta\tau$ distribution and/or $V_r$ evolution in observations. As a compromise, I will adopt the following conventions during the arguments, unless mentioned otherwise.

First, $\Delta\tau$ is taken as the stress drop during the passage of a single rupture front, known as dynamic stress drop (Kaneko and Shearer, 2015; Svetlizky et al., 2019; Ding XT et al., 2024). Second, the averaged values of $\Delta\tau$ and $V_r$ (in terms of arithmetic mean) over each considered rupture event or rupture episode will be used for comparing the results, to reduce the influence of local anomalies and estimation uncertainty; after such averaging operation, $G_s(\Delta\tau)$ can show a more explicit scaling relation with $\Delta\tau$ as $G_s(\Delta\tau) \propto (\Delta\tau)^2$, regardless of the detailed form of $G_s(\Delta\tau)$ and the exact rupture mode or type (Freund, 1998; Nielsen and Madariaga, 2003; Weng HH and Ampuero, 2019; Dong P et al., 2023). Third, $G_c$ is taken as a local property near the rupture front; when multiple rupture fronts arise from a multi-stage weakening process, separate $G_c$ will be considered near each rupture front (Ding XT et al., 2024; Paglialunga et al., 2022), rather than attributing the integrated (nonlocal) $G_c$ to a "single", smeared-out rupture front (Bolotskaya et al., 2025). The above conventions, although somewhat arbitrary and simplified, can help grasp the general essence of the arguments to be presented in Section 3, without the need to consider more realistic but non-essential complexities (Ben-Zion, 2017). Nonetheless, some of the complexities and other factors that are ignored during the arguments will be briefly discussed in Section 4.

## 3. Energy-based arguments and supporting observations

### 3.1. With constant or mildly varying fracture energy

To begin with, I consider the simplest and also the most fundamental scenario where fracture energy $G_{\mathrm{c}}$ remains constant. Since an increase in $\Delta\tau$ generally would elevate $G_{\mathrm{s}}(\Delta\tau)$ (see Eq. (1), or the scaling relation $G_{\mathrm{s}}(\Delta\tau) \propto (\Delta\tau)^2$ with $\Delta\tau$ being the averaged stress drop), to still keep a balance between $G_{\mathrm{c}}$ and $G_{\mathrm{d}}(V_{\mathrm{r}}, \Delta\tau) = g(V_{\mathrm{r}}) \cdot G_{\mathrm{s}}(\Delta\tau)$, the function $g(V_{\mathrm{r}})$ must decrease, which then would lead to an increase in $V_{\mathrm{r}}$ (Figure 1a). This argument naturally explains the positive correlation between $\Delta\tau$ ($[0.19, 0.65]$ MPa) and $V_{\mathrm{r}}$ ($[0.10, 0.99]$ $C_{\mathrm{R}}$, $C_{\mathrm{R}}$ being the Rayleigh wave speed) observed in some experimental studies, where near constant $G_{\mathrm{c}}$ ($\sim 2.5\ \mathrm{J/m^2}$) has been inferred (see Figure 3a and Figure S4a in Svetlizky et al. (2017); note the last event at the top is ignored, since its local $V_{\mathrm{r}}$ can exceed $C_{\mathrm{R}}$). In the case that $G_{\mathrm{c}}$ mildly fluctuates from one event to another ($[0.388, 0.634]\ \mathrm{J/m^2}$), a positive correlation between $\Delta\tau$ ($[0.17, 0.35]$ MPa) and $V_{\mathrm{r}}$ ($[1848, 3273]$ m/s) may still hold, provided that the aforementioned compensation between $g(V_{\mathrm{r}})$ and $G_{\mathrm{s}}(\Delta\tau)$ with opposite trend changes can keep in pace with $G_{\mathrm{c}}$ (see Table 1, Figure 2d and Figure 6a in Xu SQ et al. (2019a) and choose the "West" group; note the first event at the bottom is ignored, since the related data do not provide enough spatial coverage). Here, the averaged values of $\Delta\tau$ and $V_{\mathrm{r}}$ over a fixed rupture length or interval have been used for comparing different events in the cited studies ($[50, 150]$ mm for Svetlizky et al. (2017) and $[525, 775]$ mm for Xu SQ et al. (2019a)), following the convention mentioned in Section 2. Alternatively, using the averaged (or integrated) value of $\Delta\tau$ and the local value of $V_{\mathrm{r}}$ can also work for the results in Xu SQ et al. (2019a), as $\Delta\tau$ there shows a near uniform distribution and $V_{\mathrm{r}}$ monotonically increases with rupture length. The above results are obtained from the main rupture events under different levels of shear loading and correspond to scenario #1 in Table 1.

**Table 1**
Representative scenarios for the correlation between stress drop and rupture speed

| No. | Scenario description | $G_c$ | $g(V_r)$ | $G_s(\Delta\tau)$ | $V_r$ | $\Delta\tau$ |
|---|---|---|---|---|---|---|
| #1 | Main ruptures under near constant $G_c$ (Svetlizky et al., 2017; Xu SQ et al., 2019a), comparison among rupture events under different levels of shear loading | → Near constant or with mild variation | ↘ Decreases with increasing $V_r$ | ↗ Increases with increasing $\Delta\tau$ | ↗ Increases | ↗ Increases |
| #2 | Substantially reduced $G_c$ for a secondary rupture shortly after a primary rupture due to time-dependent healing for restoring $G_c$ (Kammer and McLaskey, 2019; Xu SQ et al., 2019b), comparison among rupture episodes over the same fault segment | ↘↘ Decreases disproportionally faster than $G_s(\Delta\tau)$ | ↘ Decreases with increasing $V_r$ | ↘ Decreases with decreasing $\Delta\tau$ | ↗ Increases | ↘ Decreases |
| #3 | Substantially increased $G_c$ due to extensive off-fault damage under large $\Delta\tau$ (Gabriel et al., 2013), comparison among rupture events under different levels of $\Delta\tau$ | ↗↗ Increases disproportionally faster than $G_s(\Delta\tau)$ | ↗ Increases with decreasing $V_r$ | ↗ Increases with increasing $\Delta\tau$ | ↘ Decreases | ↗ Increases |

*Note*. →: near constant or with mild variation; ↘ (↗): decreases (increases); ↘↘ (↗↗): substantially decreases (increases).

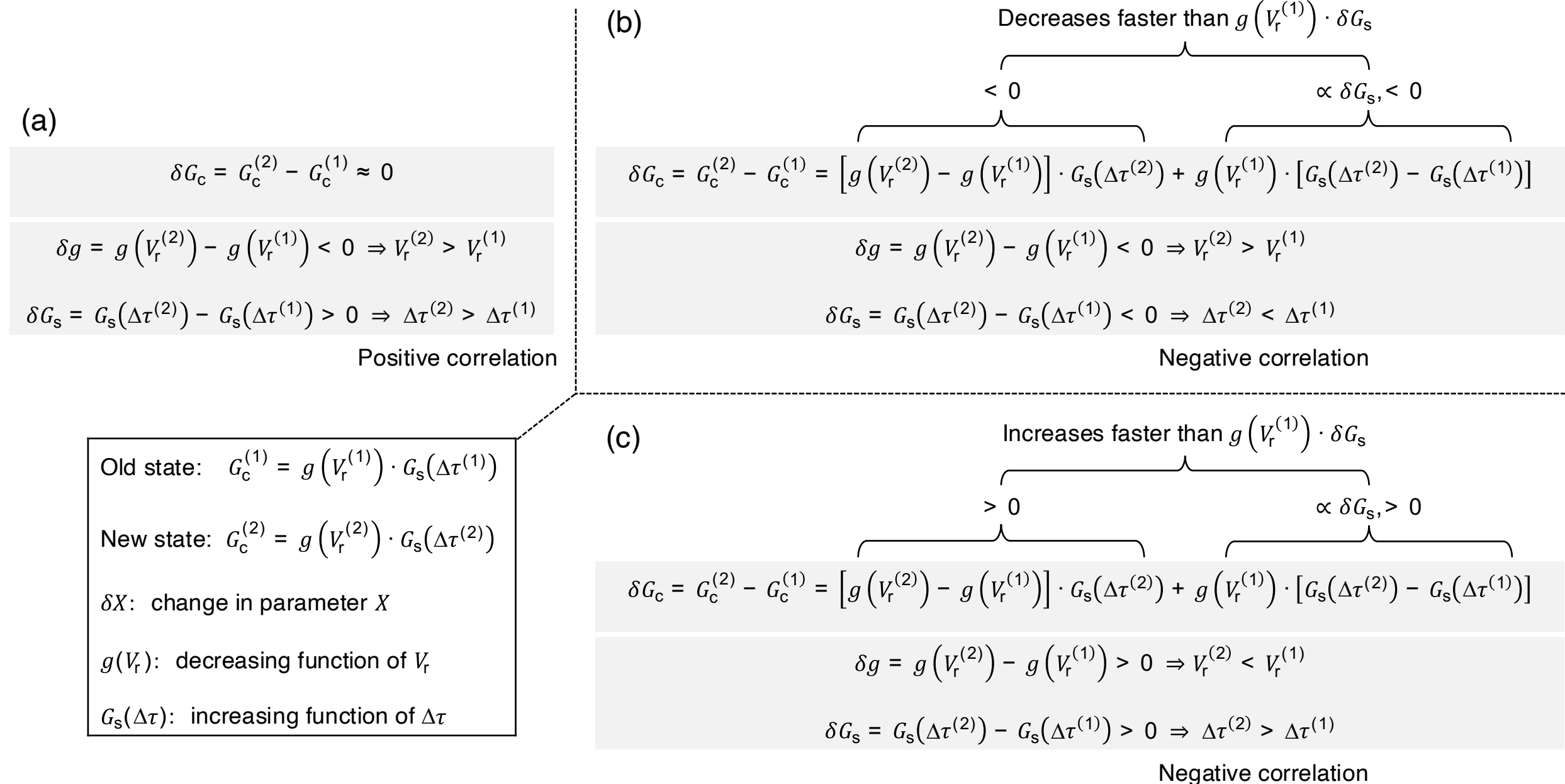

**Figure 1.** Basic derivations showing how a positive or negative correlation between stress drop $\Delta\tau$ and rupture speed $V_r$ can emerge, depending on the condition of fracture energy $G_c$. All the symbols retain their original meanings as defined in Section 2, with superscript (1) and (2) indicating an old and new state, respectively. The two states may represent two rupture events or two rupture episodes, which are used to evaluate the changes in $V_r$ and $\Delta\tau$, and then the related correlation. Panels (a), (b) and (c) correspond to scenarios #1, #2 and #3 in Table 1, respectively. Besides the presented mathematical relations, the three scenarios can also be understood by the following descriptions. Rupture speed $V_r$ is controlled by both energy supply $G_s(\Delta\tau)$ and energy consumption $G_c$, which are balanced via a proportionality factor $g(V_r)$. For (a), increasing energy supply while keeping energy consumption roughly fixed can cause an increase in $V_r$. For (b), reducing energy consumption at a rate disproportionally faster than energy supply may cause an increase in $V_r$. For (c), increasing energy consumption at a rate disproportionally faster than energy supply may cause a decrease in $V_r$.

### 3.2. With space- or time-dependent fracture energy

Next, I further relax the constraint on fracture energy $G_c$. This is motivated by several lines of evidence indicating that $G_c$ needs not to always remain constant and sometimes can vary substantially. For instance, $G_c$ can show a spatial variation by one order of magnitude ($5 - 45\ \mathrm{J/m^2}$) on a meter-scale laboratory fault, depending on local normal stress and other factors (Wang LF et al., 2024). Spatial variation in $G_c$ is thought to also exist on natural faults, as inferred by the location, size, and connectivity of asperities/barriers (Aki, 1984; Lay et al., 2012; Li JX et al., 2023). Such spatially variable $G_c$, as implied by abrupt changes in $V_r$ (see point (ii) in Section

2) and contact area over different intervals of an intermittent rupture process, can sometimes cause a negative correlation between $\Delta\tau$ (in terms of contact area reduction) and $V_\mathrm{r}$ (Rubinstein et al., 2004). In this case, increased $G_\mathrm{c}$ in the central interval appears to overtake the effect of $\Delta\tau$ in reducing $V_\mathrm{r}$: from sub-Rayleigh front of $\sim 800\ \mathrm{m/s}$ (with normalized $\Delta\tau$ of $\sim 10\%$) in the neighboring interval to slow front of $\sim 50\ \mathrm{m/s}$ (with normalized $\Delta\tau$ of $\sim 20\%$) in the central interval (see Figure 4 and Figure 5 in Rubinstein et al. (2004)). Here, however, $\Delta\tau$ and $V_\mathrm{r}$ are estimated by the averaged values over each interval (Rubinstein et al., 2004), such that one cannot completely isolate the influence of $\Delta\tau$ from non-overlapping intervals (see point (iii) in Section 2).

$G_\mathrm{c}$ can vary with time as well, e.g., by two orders of magnitude ($1$ vs. $0.01\ \mathrm{J/m^2}$) between the primary and secondary slip fronts that sequentially sweep over the same fault segment (Kammer and McLaskey, 2019). Such reduction in $G_\mathrm{c}$ can sometimes produce rapid secondary slip fronts with very low $\Delta\tau$, in contrast to their predecessors with relatively slow $V_\mathrm{r}$ and finite $\Delta\tau$ (Dunham et al., 2003; Xu SQ et al., 2019b; Guérin-Marthe, 2019; Shi SL et al., 2023; Ding XT et al., 2024; Latour et al., 2024), for instance, $V_\mathrm{r} = 3216\ \mathrm{m/s}$ and $\Delta\tau = 0.07\ \mathrm{MPa}$ (for a secondary front) vs. $V_\mathrm{r} = 1250\ \mathrm{m/s}$ and $\Delta\tau = 0.22\ \mathrm{MPa}$ (for a primary front) (see Figure 3a in Xu SQ et al. (2019b), averaged over strain gauges $\#10 - \#18$). Taken to the extreme, certain secondary slip front can propagate precisely at the Rayleigh wave speed $C_\mathrm{R}$ along the mode-II direction and the shear wave speed $C_\mathrm{S}$ along the mode-III direction (Dunham et al., 2003), and is associated with essentially zero $G_\mathrm{c}$ and zero $\Delta\tau$. Such results have important implications for the extreme form of secondary slip front. First, the corresponding $g(V_\mathrm{r})$ cannot take other values but also zero, since the propagation speed is precisely at the limiting speed ($V_\mathrm{r}^\mathrm{lim} = C_\mathrm{R}$ or $C_\mathrm{S}$) (Freund, 1998). Second, $G_\mathrm{c}$ is a higher-order zero compared to $G_\mathrm{s}(\Delta\tau)$, according to the energy balance relation $0 = G_\mathrm{c} = g\left(V_\mathrm{r}^\mathrm{lim}\right) \cdot G_\mathrm{s}(0) = 0 \cdot 0$. Together, they imply that, when considering the transition from a primary slip front to a secondary one, the decrease in $G_\mathrm{c}$ can be disproportionally faster than that in $G_\mathrm{s}(\Delta\tau)$, due to a simultaneous decrease in $g(V_\mathrm{r})$ (e.g., from a finite value towards zero, see Figure 1b).

For the above cases with recurring slip episodes, the time duration for a specific fault segment to remain locked since the last slip episode, which is related to fault healing (Ampuero and Rubin, 2008), appears to control the effective value of $G_c$ for the next slip episode (Sirorattanakul et al., 2025). This anticipation has been confirmed by observation-calibrated modeling results, based on the rate- and state-dependent friction law with an aging law for the evolution of state variable (Wang LF et al., 2024). Therefore, one can come up with an idea that the passage of the primary slip front resets $G_c$ to a low (possibly down to zero) value, and a short elapsed time afterwards (until the arrival of the secondary slip front) does not allow for a sufficient recovery of $G_c$ (Xu SQ et al., 2019b). Then, the substantially reduced $G_c$ for the secondary slip front can be balanced by a decrease in both $g(V_r)$ and $G_s(\Delta\tau)$ relative to the values for the primary slip front (which itself is preceded by a long healing time), causing a negative correlation between $\Delta\tau$ and $V_r$ (scenario #2 in Table 1, Figure 1b). To make this mechanism work, the reduction in $G_c$ must disproportionally outpace the decrease in $G_s(\Delta\tau)$ (Figure 1b), as demonstrated by the example mentioned in the preceding paragraph. Apart from the theoretical feasibility, the above scenario has been verified by several experimental observations (McLaskey et al., 2015; Kammer and McLaskey, 2019; Xu SQ et al., 2019b). It is noteworthy that a similar idea, by invoking time-dependent $G_c$ for recurring slip episodes, may explain the behaviors of slow earthquakes in the Cascadia subduction zone (Houston et al., 2011; Hawthorne et al., 2016; Bletery et al., 2017), assuming that fracture mechanics and the related ways of estimating source parameters (originally for regular, fast earthquakes) are still applicable. On the other hand, it is possible that other mechanisms mediated by slip, loading rate or specific microphysical processes may also work, as long as they can influence $G_c$ in a similar way as described above.

### 3.3. With damage-dependent fracture energy

Last but not the least, there are other ways for reaching a negative correlation between $\Delta\tau$ and $V_r$. For instance, one can extrapolate the classical concept of fracture energy $G_c$ to any dissipative processes (e.g., off-fault damage) that can effectively impede the acceleration of rupture front (Andrews, 2005; Templeton, 2009; Gabriel et al., 2024; Nielsen, 2017; Ben-Zion and Dresen, 2022; Cocco et al., 2023). For convenience, I assume that fracture energy dissipation, which now can extend to off-fault region due to damage, still occurs near the rupture front, following point (i) and the convention mentioned in Section 2. Then I consider a scenario where the entire on- and off-fault regions are on the verge of failure and then a dynamic rupture is activated along the fault. On one hand, larger $\Delta\tau$ generally favors larger $G_s(\Delta\tau)$ and hence faster $V_r$, as already explained in Section 3.1. On the other hand, larger $\Delta\tau$ can also induce more extensive off-fault damage through the enhanced stress field surrounding the rupture front (e.g., see Figure 4.6 in Templeton (2009)), which in turn can damp the acceleration of rupture front. If the damage-related increase in $G_c$ disproportionally outpaces the gain in $G_s(\Delta\tau)$, then an increase in $g(V_r)$ must be involved to keep the energy balance, causing a negative correlation between $\Delta\tau$ and $V_r$ (scenario #3 in Table 1, Figure 1c). The feasibility of this scenario has been verified by numerical simulations (Gabriel et al., 2013), especially with a high angle $\Psi$ of maximum compressive stress to the fault (e.g., $\Psi \geq 50°$). For instance, with $\Psi = 60°$, increasing the nondimensional $\Delta\tau$ from $0.25$ to $0.50$ (corresponding to a decrease of the relative strength parameter $S$ from $3.00$ to $1.00$) could reduce $V_r/C_S$ ($C_S$ being the shear wave speed) from $0.78$ to $0.50$ (see Table 1 and Figure 8 in Gabriel et al. (2013)). The above damage-related mechanism has been invoked as a possible interpretation for the inferred results on some natural earthquakes (Chounet et al., 2018; Nazeri et al., 2025; Žilić et al., 2025), despite the difficulties in robustly constraining the related source parameters (see the second paragraph in Section 1). The damage-related mechanism may also explain why some aseismic (slow) slip events are associated with quite large stress drops (Luo H et al., 2025), assuming that fracture mechanics and the related analysis method can still be applied to such events.

## 4. Summary and outlook

On the basis of the energy balance relation from fracture mechanics, I have shown that the correlation between stress drop $\Delta\tau$ and rupture speed $V_r$ is not necessarily fixed, but can change with the condition of fracture energy $G_c$. Using the averaged values of $\Delta\tau$ and $V_r$ (in terms of arithmetic mean) over each rupture event or episode for comparison, theoretical arguments and collected observational results have confirmed that near constant $G_c$ can yield a positive correlation between $\Delta\tau$ and $V_r$, whereas variable $G_c$ dependent on space, time or off-fault damage may cause a negative correlation between $\Delta\tau$ and $V_r$. This flexible correlation between $\Delta\tau$ and $V_r$ resembles the famous example in Physics 101, where electric power $P$ can positively correlate with resistance $R$ if current $I$ is fixed (i.e., $P = I^2 \cdot R$), but can also negatively correlate with resistance $R$ if voltage $V$ is fixed (equivalently, if current $I$ is allowed to vary) (i.e., $P = V^2/R$). The take-home message is that one may need to examine the condition of other factors (e.g., $G_c$) when evaluating the correlation between the parameters of concern (e.g., $\Delta\tau$ and $V_r$). This message may seem straightforward from the perspective of a high-dimensional parameter space and the physical principles of the problem, but has not been paid enough attention by the seismology community, where correlation is often explored empirically between two sets of parameters without checking the condition of other factors (Causse and Song, 2015; Chounet et al., 2018; Shimmoto, 2022; Nazeri et al., 2025).

It should be noted that the above conclusion on a flexible correlation between $\Delta\tau$ and $V_r$ does not significantly rely on the choice of using arithmetic mean for the examined parameters, which merely represents a compromise to balance theoretical considerations and available observations. Especially, in several studies equipped with near-field observations (Xu SQ et al., 2019a and 2019b; Gabriel et al., 2013), the local $\Delta\tau$ shows a near uniform distribution while the local $V_r$ either monotonically increases or remains roughly constant. At least for these basic cases with simple distribution and evolution of source parameters (which themselves are well constrained), a consistent correlation (e.g., positive for the data in Xu SQ et al. (2019a), negative for the data in Gabriel et al. (2013) under a high $\Psi$) could still hold among different rupture events, regardless of the detailed way of computing $\Delta\tau$ and $V_r$, including the one of using integrated $\Delta\tau$ and local $V_r$ as in Eq. (1). Clearly, these basic cases can serve as simple counter example to each other, which is sufficient to refute a fixed (always positive or negative) correlation between $\Delta\tau$ and $V_r$. On the other hand, for other cases especially those with strong heterogeneities, different averaging procedures may lead to different values for $\Delta\tau$ and $V_r$, which can further affect the deduced correlation between $\Delta\tau$ and $V_r$ or the application of fracture mechanics. Moreover, the definition of rupture itself can be tricky in such cases, because there may exist multiple rupture fronts or subevents (Danré et al., 2019; Ding XT et al., 2024). Therefore, caution must be taken for analyzing the cases with more complicated distribution or evolution of source parameters.

Before closing this article, I would like to outline the following issues for further investigation. First, the same idea by using the energy balance relation may also suggest a weak or even null correlation between $\Delta\tau$ and $V_r$, if the change in $G_c$ is almost perfectly matched by the change in $G_s(\Delta\tau)$ (implying little or no change in $V_r$). It will be interesting to explore whether such scenario can be realized in actual observations. Second, the way to estimate $\Delta\tau$ and $V_r$ may vary between different studies: $\Delta\tau$ and $V_r$ can be directly and independently measured near the fault for laboratory earthquakes (Svetlizky et al., 2017; Xu SQ et al., 2019a), while are typically inferred from far-field observations for natural earthquakes (Chounet et al., 2018). Moreover, the estimations of $\Delta\tau$ and $V_r$ for natural earthquakes are often coupled (i.e., not independent) through the scaling relation $\Delta\tau \cdot (V_r)^3 \propto M_0$ ($M_0$ denotes seismic moment) and thus can be subject to uncertainty and tradeoff issues (Kanamori and Rivera, 2004; Abercrombie, 2021; Ye LL et al., 2016). Third, even for the same study, one has to distinguish between dynamic and static stress drops (Kanamori and Rivera, 2006; Passelègue et al., 2016), crack-type and pulse-type ruptures (Lambert et al., 2021), and the local and average values of rupture speed or stress drop especially for cases with strong heterogeneities (Cheng C et al., 2023; Ding XT et al., 2025; Noda et al., 2013; Xu LW et al., 2023). Finally, when additional phases or stress drops occur at later stages after the passage of the primary rupture front, one needs to judge whether they can still influence the energy balance for the primary rupture front. Although those phases or stress drops were traditionally assumed to contribute to a nonlocal form of $G_c$ (also called the breakdown work) (Abercrombie and Rice, 2005; Cocco et al., 2023), they are not necessarily relevant to the propagation of the primary rupture front. For instance, they may reflect a healing process during which the primary rupture front has already stopped (Madariaga, 1976; Kaneko and Shearer, 2015; Ke CY et al., 2022), or their associated propagating fronts may not catch up with the primary rupture front within the considered space and time (Wada and Goto, 2012; Ding XT et al., 2024). The above and other related issues should be investigated by the follow-up works, to further enhance the understanding of stress drop and rupture speed, as well as the contribution to earthquake physics and seismic hazard assessment.

**Acknowledgments**

I thank Yehuda Ben-Zion and Rachel Abercrombie for useful suggestions, two anonymous reviewers for providing comments, and the editors for handling the manuscript. I also thank Eiichi Fukuyama and Futoshi Yamashita for guiding me to the field of laboratory earthquakes, and Lingling Ye for fruitful discussion on the observation of natural earthquakes. I am grateful to Xiaotian Ding and Jun Xie for the assistance in reading the data from some observational studies. This work was supported by National Natural Science Foundation of China (42474082, 42074048) and Guangdong Program (2021QN02G106).

## Data availability

The mentioned values of source parameters (fracture energy, stress drop, rupture speed) can be found or derived from previously published studies, including Dunham et al. (2003), Gabriel et al. (2013), Kammer and McLaskey (2019), Rubinstein et al. (2004), Svetlizky et al. (2017), Wang LF et al. (2024), Xu SQ et al. (2019a, 2019b).

## Conflict of interest

The author affirms that he has no financial and personal relationships with any individuals or organization that could have potentially influenced the work presented in this paper.